\documentclass[prb,
twocolumn,
superscriptaddress,showpacs,amsmath,amssymb]{revtex4}

\begin{document}

\author{G.E.~Volovik}
\affiliation{Low Temperature Laboratory, Aalto University,  P.O. Box 15100, FI-00076 Aalto, Finland}
\affiliation{Landau Institute for Theoretical Physics, acad. Semyonov av., 1a, 142432,
Chernogolovka, Russia}

\title{Particle creation: Schwinger + Unruh + Hawking}

\date{\today}

\begin{abstract}
{We discuss the interconnection between the Schwinger pair creation in electric field, Hawking radiation and particle creation in the Unruh effect. All three processes can be described in terms of the entropy and temperature. These thermodynamic like processes can be combined. We consider the combined process of creation of charged and electrically neutral particles in the electric field, which combine the Schwinger  and Unruh effects. We also consider the creation of the charged black and white  holes in electric field, which combines the Schwinger effect and the black hole entropy. The combined processes obey the sum rules for the entropy and for the inverse temperature. Some contributions to the entropy and to the temperature are negative, which reflects the quantum entanglement between the created objects.
}
\end{abstract}
\pacs{
}

\maketitle 
 
  \section{Introduction}

In this paper  we discuss the connection of Schwinger particle creation in the constant electric field\cite{Schwinger1951,Schwinger1953} with the particle production in the Unruh\cite{Unruh1976} and Hawking\cite{Hawking1974} effects. For that we consider the combined effects, which involve simultaneously the Schwinger particle production and the other effects. These  combined effects demonstrate that the entropy and temperature can be associated not only with the event horizons, as it was suggested  by Gary Gibbons and Stephen Hawking,\cite{GibbonsHawking1977} but also can be extended to the Schwinger effect.

The plan of the paper is the following. In Sec. \ref{SchwingerSec} we recall the derivation of the Schwinger pair production.

In Sec. \ref{dSvsSchwinger} the combined Schwinger and Unruh process is considered. It is the correlated process of creation of particle with mass $M+m$ in electric field, in which the charged particle with mass $M$ is created in the  Schwinger process, and chargeless particle with mass $m\ll M$ is created in the  Unruh process, which accompanies the Schwinger creation.
This correlated Schwinger + Unruh process suggests that the Schwinger pair creation can be also characterized by the temperature, and thus by the corresponding entropy, which is discussed in Sec.\ref{Entropy}. For the Schwinger pair creation the entropy appears to be negative, which reflects the quantum entanglement of the created particles.

In Sec. \ref{Ssum} the combined Schwinger + Hawking process is considered, in which the pair of the charged black holes is created in external electric field. It is shown that this processes obeys the sum rule for the entropy, i.e. the total entropy is the sum of the Schwinger negative entropy and positive entropies of the created black holes. The same sum rule is applicable to creation of the pair of charged white holes, and to creation of black hole - white hole paper. The similar sum rule takes place  for the inverse temperature.

Here we do not consider the Schwinger pair production in curved spacetime (see e.g. Ref. \cite{Dyon2018} and references therein, and also the analog system, in which both gravity and electromagnetic field are emergent effective fields\cite{Schopohl1992}). Although this also combines Schwinger, Unruh and Hawking processes, they do not reflect the proper thermodynamics.

 \section{Schwinger process in semiclassical approximation}
 \label{SchwingerSec}

  \subsection{Quantum tunneling approximation}
 \label{QuantumTunnelingSec}

The spectrum of a charged particle in a constant electric field:
\begin{equation}
E({\bf p},{\bf r})= \pm \sqrt{M^2 +p^2} + q{\bf {\cal E}}\cdot {\bf r} \,,
\label{Schwinger}
\end{equation}
where $q$ is electric charge.
Introducing the coordinate  $z$ along the electric field, one obtains for Schwinger case:
\begin{equation}
E({\bf p},z)= \pm \sqrt{M^2 +p_\perp^2+p_z^2} + q {\cal E} z \,.
\label{SchwingerZ}
\end{equation}

 In the semiclassical approximation the probability  of the particle creation is given by the tunneling exponent $2{\rm Im}\int dz p_z(z)$:
 \begin{equation}
W=\sum_{\bf p}w_{\bf p}=\sum_{\bf p}\exp{\left(- 2{\rm Im}\int dz p_z(z) \right)}\,.
\label{Probability}
\end{equation}
where the tunneling trajectories along $z$ are given by equation $E({\bf p},z)=E$.
In case of Schwinger pair production the tunneling exponent  depends only on the transverse momentum ${\bf p}_\perp$:
\begin{equation}
w_{\bf p}^{\rm Schw}=\exp{\left(- \frac{\pi \tilde M^2}{q\cal E} \right)}\,\, , \,\, \tilde M^2=M^2 +p_\perp^2 \,,
\label{SchwingerExp}
\end{equation}
The semiclassical approximation is valid for $M^2 \gg q{\cal E}$.

  Integration over transverse momenta gives
\begin{equation}
\int \frac{d^2 p_\perp}{(2\pi)^2}\exp{\left(- \frac{\pi \tilde M^2}{q\cal E} \right)}= 
 \frac{q\cal E}{(2\pi)^2}  \exp{\left(- \frac{\pi  M^2}{q\cal E} \right)} \,.
\label{SchwingerExpIntegrate}
\end{equation}
Integral over $dp_z /2\pi$ diverges because the exponent does not depend on $p_z$. Due to the motion equation  $dp_z = q{\cal E}dt$, one obtains the known Schwinger pair creation per unit volume per unit time (the integer factors for polarization and for spin of particles are ignored):
\begin{equation}
\Gamma^{\rm Schw}(M) = \frac{dW^{\rm Schw}}{dt} = 
 \frac{q^2{\cal E}^2}{(2\pi)^3}  \exp{\left(- \frac{\pi  M^2}{q\cal E} \right)} \,.
\label{SchwingePairCreation}
\end{equation}

\subsection{Comparison with Unruh radiation and the factor 2 problem}
 \label{Factor2}

The straightforward  comparison of Eq.(\ref{SchwingePairCreation}) with the Unruh radiation in the accelerated frame reveals the factor of 2 problem. The Unruh temperature is $T_{\rm U}=a/2\pi$, where $a$ is acceleration. Since in the Schwinger case the  acceleration in the electric field is $a=q{\cal E}/M$, one obtains:
\begin{equation}
 \exp{\left(- \frac{\pi  M^2}{q\cal E} \right)} = \exp{\left(- \frac{\pi  M}{a} \right)}= \exp{\left(- \frac{M}{2T_{\rm U}} \right)}\,.
\label{factor2discrepancy}
\end{equation}
In this naive approach the temperature of thermal radiation is twice the Unruh temperature. The factor of 2 problem arises also in the consideration of the Hawking radiation in the de Sitter expansion, see Ref. \cite{Volovik2022b} and references therein.
Subtleties in the tunneling approach to Hawking and Unruh radiation see in Refs. 
\cite{Akhmedov2006,Akhmedov2007,Akhmedov2008,Akhmedova2008,Akhmedova2009,Gill2010}.
Different scenarios of resolving the above discrepancy between the Schwinger and Unruh mechanisms see in Refs.\cite{Parentani1997,KharzeevTuchin2005,Zabrodin2022} and references therein. We consider the scenario somewhat similar to that in Ref.\cite{KharzeevTuchin2005}.

 \section{Correlated Schwinger + Unruh process}
 \label{dSvsSchwinger}
 
From Eq.(\ref{SchwingePairCreation}) it follows that the probability of creation of particle with mass $M+m$ and charge $q$  in the limit $m\ll M$ can be expressed in terms of the probability of creation of particle with mass $M$ and the extra term:
\begin{eqnarray}
\Gamma^{\rm Schw}(M+m) =\Gamma^{\rm Schw}(M)   \exp{\left(- \frac{2\pi  Mm}{q\cal E} \right)}\,.
\label{Mm}
\end{eqnarray}
The extra term can be described  in terms of the temperature of the Unruh radiation in the accelerated frame: 
\begin{eqnarray}
  \exp{\left(- \frac{2\pi  Mm}{q\cal E} \right)}=   \exp{\left(- \frac{m}{T_{\rm U}} \right)} = \Gamma^{\rm Unruh}(m)\,,
 \label{Unruh}
 \\
T_{\rm U} = \frac{a}{2\pi} \,\,,\,\, a=\frac{q\cal E}{M}\,,
 \label{UnruhT}
  \end{eqnarray}
  where $a$ as before is the acceleration of charged particle with mass $M$ in electric field.
  
Altogether this gives
\begin{eqnarray}
\Gamma^{\rm Schw}(M+m) =\Gamma^{\rm Schw}(M)  \, \Gamma^{\rm Unruh}(m) \,,
 \label{SchwingerUnruh}
 \end{eqnarray}
 This Eq.(\ref{SchwingerUnruh}) has the following interpretation: the process on the left side of Eq.(\ref{SchwingerUnruh}) can be considered as the combination of two correlated (entangled) processes on the right side of Eq.(\ref{SchwingerUnruh}). The first process is the Schwinger creation of two particles with large mass $M$ and the charges $\pm q$. The second term describes the creation of two neutral particles, $q=0$, with masses $m\ll M$. These electrically neutral particles do not feel the electric field. But each neutral particle is entangled with its charged partner, moves along the same trajectory and thus  feels the same acceleration $a=\pm q{\cal E}/M$. As a result, the creation of the neutral particles is fully described by the Unruh process.

 This coherent combined process can be also interpreted in the following way. The charged particle with heavy mass $M$ is created by Schwinger process and moves with the acceleration $a=q{\cal E}/M$. In the accelerated reference frame (in  Rindler spacetime) the moving massive particle plays the role of the detector in the Unruh vacuum. This detector experiences the emission of a neutral particles  -- the Unruh radiation with the Unruh temperature $T_{\rm U} = a/2\pi$. 
 
 The combination of the several processes, which gives rise to the product of probabilities,  is similar to the phenomenon of cotunneling in the electronic systems. In these condensed matter systems, the electron experiences the coherent sequence of tunneling events:  from an initial to the virtual intermediate states and then to the final state.\cite{Feigelman2005,Glazman2005} In our case the virtual intermediate state is the state of the created charged particle with mass $M$. Its motion with acceleration $a$ triggers the creation of the neutral particle by the Unruh mechanism.

The combination of two processes in Eq.(\ref{SchwingerUnruh}) -- the Schwinger creation of mass $M$ and the Unruh creation of mass $m\ll M$ --  is similar to the combination of two processes in the creation of pairs of Reissner-Nordstr\"om (RN) monopole black holes  in magnetic field:\cite{Garfinkle1994} 
\begin{eqnarray}
\Gamma^{\rm BH,Monopole} =\Gamma^{\rm Monopole} \, \Gamma^{\rm BH}\,,
\label{MHmonopole}
\\
 \Gamma^{\rm BH} =\exp{\left(S_{\rm BH}\right)} \,.
  \label{BH}
  \end{eqnarray}
According to Ref. \cite{Garfinkle1994}, the instanton amplitude contains an explicit factor corresponding to the black hole entropy, $S_{\rm BH}=A/4$ in Eq.(\ref{BH}), where $A$ is the area of the event horizon. The Schwinger pair creation is modified by the black hole entropy and thus by the Hawking temperature.

We consider such combination in Sec. \ref{Ssum}.

 \section{Quantum tunneling and entropy}
 \label{Entropy}
 
In principle, the black hole thermodynamics can be obtained by consideration of the tunneling processes.  The Hawking temperature can be derived by comparing the tunneling rate with the Boltzmann factor.\cite{Volovik1999,Srinivasan1999,Wilczek2000} The black hole entropy can be found\cite{Parikh2006,Volovik2021} by comparing the tunneling rate with the thermodynamic fluctuations.\cite{Landau_Lifshitz} In this approach the tunneling process is considered as fluctuation, with the 
 probability of fluctuation being determined by the entropy difference between the initial and final states, $w \propto \exp{\left(S_{\rm final} -S_{\rm initial}\right)}$. Let us consider the process of Schwinger pair creation as the thermodynamic fluctuations, which can be expressed via the corresponding entropy $S_{\rm Schw}(M)$.
 
 \subsection{First guess}
 \label{guess}
 
 At first glance the Eq.(\ref{SchwingePairCreation}) can be written as:
 \begin{equation}
\Gamma^{\rm Schw}(M)=   \frac{q^2{\cal E}^2}{(2\pi)^3} \exp{\left(- S_{\rm Schw}(M) \right)}\,.
\label{SchwEntropy}
\end{equation}
This kind of entropy may give the Unruh temperature:
  \begin{equation}
T_{\rm Schw}^{-1}=\frac{dS_{\rm Schw}}{dM}=\frac{2\pi M}{q\cal E} =\frac{2\pi}{a}=T_{\rm U}^{-1} \,.
\label{UnruhTemp}
\end{equation}
However, this guess is not correct. The situation is different and is more interesting, see Sec.\ref{NegativeEntropy} below.

 \subsection{Negative entropy}
 \label{NegativeEntropy}

In the entropy-fluctuation relation\cite{Landau_Lifshitz,Parikh2006} the probability of fluctuation is $w \propto \exp{\left(S_{\rm final} -S_{\rm initial}\right)}$. The initial state is the vacuum, and its entropy is zero, $S_{\rm initial}=S_{\rm vac}=0$. Thus the Schwinger  entropy is negative, $S_{\rm final}\equiv S_{\rm Schw}(M)=- \frac{\pi  M^2}{q\cal E}<0$, and 
\begin{equation}
\Gamma^{\rm Schw}(M)=   \frac{q^2{\cal E}^2}{(2\pi)^3} \exp{\left(S_{\rm Schw}(M) \right)}\,.
\label{SchwEntropy}
\end{equation}
The negative entropy comes from the quantum entanglement between the created particles and also between the particles and quantum fields in the vacuum, see discussions of correlations between the radiated particles in Refs. \cite{BaochengZhang2009,Kharzeev2021}.

Then the corresponding Schwinger  temperature is also negative and is opposite to the Unruh temperature, $T_{\rm Schw}=-T_{\rm Unruh}$:
  \begin{equation}
T_{\rm Schw}^{-1}=\frac{dS_{\rm Schw}}{dM}=-\frac{2\pi M}{q\cal E} =-\frac{2\pi}{a}=-T_{\rm U}^{-1} \,.
\label{MinuaUnruhTemp}
\end{equation}
The negative temperature is a well defined quantity, especially in condensed matter systems, see review Ref. \cite{Baldovin2021}. 

 \section{Combined Schwinger+Hawking}
 \label{SchwingerHawking}

  \subsection{Sum rule for entropy}
 \label{Ssum}

The description of the quantum tunneling in terms of the thermodynamic fluctuation\cite{Landau_Lifshitz} can be also applied to the creation of two charged RN black holes discussed in Ref. \cite{Garfinkle1994}. This corresponds to the triple cotunneling -- the coherent sequence of three tunneling events. Adding the entropy of two created black holes one obtains for the full entropy, which determines the probability of creation:
 \begin{eqnarray}
S_{\rm final} -S_{\rm initial}=
\nonumber
\\
=S_{\rm Schw}(M,q) +S_{\rm RN\,BH}(M,q) +S_{\rm RN\,BH}(M,q) =
\label{BHpairs1}
\\
= - \frac{\pi  M^2}{q\cal E} + 4\pi GM^2 + 4\pi GM^2 \,.
\label{BHpairs2}
\end{eqnarray}
Here it is assumed that $q{\cal E} \ll G^{-1} \sim E_{\rm Planck}^2$, i.e. the force between the charges is much smaller than the Planck "maximum force" limit conjectured by Gibbons\cite{Gibbons2003} (see also recent papers\cite{Faraoni2021,Schiller2021} and references therein, and the criticism in Ref.\cite{Volovik2022}).
 As a result, the total contribution to Eq.(\ref{BHpairs2}) is negative.
We also take into account that the entropy of the RN black hole does not depend on charge $q$, which is due to the correlations between the inner and outer horizons.\cite{Volovik2021}

As in  Ref. \cite{Garfinkle1994} (see also \cite{Hawking1995} and references in \cite{Astorino2014}), the positive terms in Eq.(\ref{BHpairs2}) demonstrates that due to the positive entropy of the black holes, the creation of pairs  of Reissner-Nordstr\"om black holes is enhanced compared to pair production of particles or  magnetic monopoles with the same masses and charges.  

On the other hand the creation of the pairs of the RN white holes should be suppressed due to their negative entropy, 
$S_{\rm WH}=-S_{\rm BH}$:\cite{Volovik2021}
\begin{eqnarray}
S_{\rm final} -S_{\rm initial}=
\nonumber
\\
=S_{\rm Schw}(M,q) +S_{\rm RN\,WH}(M,q) +S_{\rm RN\,WH}(M,q)=
\label{WHpairs1}
\\
= - \frac{\pi  M^2}{q\cal E}  - 8\pi GM^2 \,.
\label{WHpairs2}
\end{eqnarray}

There can be also the process of the creation of the pair: black hole  + white hole. In this process the conventional Schwinger production is restored due to mutual cancellation of entropies of black and white holes:
\begin{eqnarray}
S_{\rm final} -S_{\rm initial}=
\nonumber
\\
= S_{\rm Schw}(M,q) +S_{\rm RN\,WH}(M,q) + S_{\rm RN\,BH}(M,q)  =
\label{WHBH}
\\
= S_{\rm Schw}(M,q) =- \frac{\pi  M^2}{q\cal E} .
\label{WHBH1}
\end{eqnarray}
 
  \subsection{Sum rule for temperature}
 \label{Tsum}

The sum rule for entropy in Eqs. (\ref{BHpairs1}), (\ref{WHpairs1}) and (\ref{WHBH}) is also valid for the inverse temperatures.
The corresponding temperature $1/T= dS_{\rm final}/dM$ describing the above three processes of creation is:
\begin{eqnarray}
\frac{1}{T}= \frac{1}{T_{\rm Schw}} + \frac{1}{T_{\rm BH}} + \frac{1}{T_{\rm BH}} \,\,, \,\, T_{\rm BH}=\frac{1}{8\pi G M}  \,,
\label{BHandBH}
\\
\frac{1}{T}= \frac{1}{T_{\rm Schw}} + \frac{1}{T_{\rm WH}} + \frac{1}{T_{\rm WH}} \,\,, \,\, T_{\rm WH}=-\frac{1}{8\pi G M}  \,,
\label{WHandWH}
\\
\frac{1}{T}= \frac{1}{T_{\rm Schw}} + \frac{1}{T_{\rm BH}} + \frac{1}{T_{\rm WH}} = \frac{1}{T_{\rm Schw}}   \,.
\label{BHandWH}
\end{eqnarray}

The summation law for inverse temperature is similar to that, which takes place for the black hole with several horizons,\cite{Volovik2021} where the temperature is expressed in terms of the Hawking temperatures on different horizons. For the black hole with two horizons, such as RN black hole,  one has $1/T_{\rm RN\, BH}= 1/T_+ +1/T_-$, where $T_+$ and $T_-$ are the  temperatures of outer and inner horizons correspondingly. The temperature $T_{\rm RN\, BH}$ determines the Hawking radiation rate, which corresponds to cotunneling --  the coherent sequence of tunneling events at different horizons: 
$\exp{\left(-E/T\right)} =\exp{\left(-E/T_-\right)} \exp{\left(-E/T_+\right)}$. 

The same was found for the Schwarzschild-de Sitter black hole,\cite{LiMaMa2017,Nakarachinda2017}  with $T_+$ and $T_-$ being the  temperatures of the cosmological horizon and the black hole horizon correspondingly.
On the other hand,  the total entropy of the RN black hole is not necessarily the sum of the entropies of the inner and outer horizons due to correlations between the  two horizons, see e.g. Refs.\cite{Volovik2021,YunHe2018,YuboMa2021,Odintsov2021,Shankaranarayanan2003,Singha2021,Azarnia2021}.  
Also the consideration of the black holes in the de Sitter environment requires the use of special modifications of the Painleve-Gullstrand coordinates,\cite{Volovik2022c} which may modify the thermodynamics.

  \section{Discussion}
 \label{DiscussinSec}

We considered the extension of the phenomenon of the black hole entropy to the general  processes of quantum tunneling, including  the macroscopic quantum tunneling. It appears that the Schwinger pair creation obeys the same thermodynamic laws as in the black hole thermodynamics. The thermodynamics naturally connects the Schwinger pair creation in electric field with the Hawking radiation and with the particle creation in the Unruh effect. All three processes can be described in terms of the entropy and temperature. 
Moreover, these 3 processes can be combined. For example, the process of creation of charged and electrically neutral particles in the electric field combines the Schwinger  and Unruh effects. The same takes place in the process of the creation of the charged black and white  holes in electric field. This process combines the Schwinger effect and the black hole entropy. The combined processes obey the sum rules for the entropy and for the inverse temperature. Some contributions to the entropy and to the inverse temperature are negative, which reflects the quantum entanglement between the created objects: charged particles, magnetic monopoles and black and white holes.

 {\bf Acknowledgements}. I thank D. Kharzeev and K. Tuchin for correspondence. This work has been supported by the European Research Council (ERC) under the European Union's Horizon 2020 research and innovation programme (Grant Agreement No. 694248).


\begin{thebibliography}{99}

\bibitem{Schwinger1951}
J. Schwinger, 
On gauge invariance and vacuum polarization,
 Phys. Rev. {\bf 82}, 664--679 (1951).
 
 \bibitem{Schwinger1953}
J. Schwinger, 
The theory of quantized fields. V,
Phys. Rev. {\bf 93},  616 (1953)

\bibitem{Unruh1976}
W.G. Unruh, 
Notes on black-hole evaporation,
Phys. Rev. D {\bf 14}, 870 (1976).

\bibitem{Hawking1974}
S.W. Hawking,
Black hole explosions?, 
Nature {\bf 248}, 30--31 (1974).

\bibitem{GibbonsHawking1977}
G.W. Gibbons and S.W. Hawking,
Cosmological event horizons, thermodynamics, and particle creation,
Phys. Rev. D {\bf 15}, 2738 (1977).

\bibitem{Dyon2018}
Chiang-Mei Chen, Sang Pyo Kim, Jia-Rui Sun and Fu-Yi Tang,
Pair production of scalar dyons in Kerr–Newman black holes,
Physics Letters B {\bf 78}, 129--138 (2018).

\bibitem{Schopohl1992}
N. Schopohl and G.E. Volovik, 
Schwinger pair production in the orbital dynamics of $^3$He-B, 
Annals of Phys. {\bf 215}, 372--385 (1992).

\bibitem{Volovik2022b}
G.E. Volovik,
Double Hawking temperature: from black hole to de Sitter,
arXiv:2205.06585 [gr-qc].

\bibitem{Akhmedov2006}
E.T. Akhmedov, V. Akhmedova and D. Singleton,
Hawking temperature in the tunneling picture,
Phys. Lett. B {\bf 642}, 124--128 (2006),

\bibitem{Akhmedov2007}
E.T. Akhmedov, V. Akhmedova, T. Pilling and D. Singleton,
Thermal radiation of various gravitational background,
Int. J. Mod. Phys. A {\bf 22} 1705 (2007). 

\bibitem{Akhmedov2008}
E.T. Akhmedov, T. Pilling and D. Singleton, 
Subtleties in the quasi-classical calculation of Hawking radiation, 
Int. J. Mod. Phys. D {\bf 17},  2453 (2008).

\bibitem{Akhmedova2008}
V. Akhmedova, T. Pilling, A. de Gill, and D. Singleton, 
Temporal contribution to gravitational WKB-like calculations,
Phys. Lett. B {\bf 666}, 269--271 (2008). 

\bibitem{Akhmedova2009}
V. Akhmedova, T. Pilling, A. de Gill, and D. Singleton, 
Comments on anomaly versus WKB/tunneling methods for calculating Unruh radiation,
 Phys. Lett. B {\bf 673}, 227--231 (2009). 

\bibitem{Gill2010}
A.  de Gill,
A WKB-like approach to Unruh radiation,
American Journal of Physics {\bf 78}, 685 (2010).

\bibitem{Parentani1997}
R. Parentani and S. Massar,
Schwinger mechanism, Unruh effect, and production of accelerated black holes,
Phys. Rev. D {\bf 55}, 3603 {1997}.

\bibitem{KharzeevTuchin2005}
D. Kharzeev and K. Tuchin,
From color glass condensate to quark–gluon plasma through the event horizon,
Nuclear Physics A {\bf 753},  316--334 (2005).

\bibitem{Zabrodin2022}
M. Teslyk, O. Teslyk, L. Zadorozhna, L. Bravina and E. Zabrodin,
Unruh effect and information entropy approach,
Particles {\bf  5}, 157--170 (2022).

\bibitem{Feigelman2005}
M.V. Feigel’man and A.S. Ioselevich,
Variable-range cotunneling and conductivity of a granular metal,
JETP Letters {\bf 81},  277--283  (2005).

\bibitem{Glazman2005}
L.I. Glazman and  M. Pustilnik,
Low-temperature transport through a quantum dot,
Lectures notes of the Les Houches Summer School 2004,
in:  "Nanophysics: Coherence and Transport," eds. H. Bouchiat et al. (Elsevier, 2005), pp. 427--478,
arXiv:cond-mat/0501007.

\bibitem{Garfinkle1994}
D. Garfinkle, S.B.Giddings and  A. Strominger,
Entropy in black hole pair production,
Phys. Rev. {\bf 49}, 958 (1994).

\bibitem{Volovik1999}
G.E. Volovik,  
Simulation of Painleve-Gullstrand black hole in thin $^3$He-A film,  
Pis'ma ZhETF {\bf 69}, 662 -- 668 (1999), JETP Lett.  {\bf 69}, 705 -- 713 (1999); 
gr-qc/9901077.

\bibitem{Srinivasan1999}
K. Srinivasan and T. Padmanabhan,
Particle production and complex path analysis,
Phys. Rev. D {\bf 60}, 024007 (1999).

\bibitem{Wilczek2000}
M.K. Parikh and F. Wilczek, 
Hawking radiation as tunneling,
Phys. Rev. Lett. {\bf 85}, 5042 (2000).

\bibitem{Parikh2006}
M.K. Parikh,
Energy Conservation and Hawking Radiation,
The Tenth Marcel Grossmann Meeting, pp. 1585--1590 (2006),
hep-th/0402166.

\bibitem{Volovik2021}
G.E. Volovik,
Macroscopic quantum tunneling: from quantum vortices to black holes and Universe,
to be published in JETP issue devoted to Rashba-95,
arXiv:2108.00419.

\bibitem{Landau_Lifshitz}
 L.D. Landau  and  E.M. Lifshitz, 
 Course of Theoretical Physics, Volume 5, 
 Statistical Physics.

\bibitem{BaochengZhang2009}
Baocheng Zhang, Qing-yu Cai, Li You and Ming-sheng Zhan,
Hidden messenger revealed in Hawking radiation: A resolution to the paradox of black hole information loss,
Physics Letters B {\bf 675}, 98--101 (2009).

\bibitem{Kharzeev2021}
A. Florio and D.E. Kharzeev,
Gibbs entropy from entanglement in electric quenches,
Phys. Rev. D {\bf 104}, 056021 (2021).

\bibitem{Baldovin2021}
M. Baldovin, S. Iubini, R. Livi and A. Vulpiani,
 Statistical mechanics of systems with negative temperature,
 Phys. Reports  {\bf 923}, 1--50 (2021),
  arXiv:2103.12572.

\bibitem{Hawking1995}
S.W. Hawking, G.T. Horowitz and S.F. Ross,
Entropy, area, and black hole pairs,
Phys. Rev. D {\bf 51}, 4302 (1995).

\bibitem{Astorino2014}
M. Astorino,
Pair creation of rotating black holes,
Phys. Rev. D {\bf 89}, 044022 (2014).

\bibitem{Gibbons2003}
G.W. Gibbons,
The maximum tension principle in general relativity,
Foundations of Physics {\bf 32}, 1891--1901 (2003).

\bibitem{Faraoni2021}
V. Faraoni,
Maximum force and cosmic censorship,
Phys. Rev. D {\bf 103}, 124010 (2021).

\bibitem{Schiller2021} 
C. Schiller,
Tests for maximum force and maximum power,
Phys. Rev. D {\bf 104}, 124079 (2021).

\bibitem{Volovik2022}
G.E. Volovik,
Negative Newton constant may destroy some conjectures,
Modern Physics Letters A  {\bf 37}, 2250034 (2022),
arXiv:2202.12743.

\bibitem{LiMaMa2017}
H. F. Li, M. S. Ma, and Y. Q. Ma, 
Thermodynamic properties of black holes in de Sitter space, 
Mod. Phys. Lett. A {\bf 32}, 1750017 (2017).

\bibitem{Nakarachinda2017}
R. Nakarachinda, E. Hirunsirisawat, L. Tannukij and P. Wongjun,
Effective thermodynamical system of Schwarzschild–de Sitter black holes from Renyi statistics,
Phys. Rev. D {\bf 104}, 064003 (2021).

\bibitem{YunHe2018}
Yun He, Meng-Sen Ma and Ren Zhao,
Entropy of black holes with multiple horizons,
Nucl. Phys. B {\bf 930}, 513--523 (2018).

\bibitem{YuboMa2021}
Yubo Ma, Yang Zhang, Lichun Zhang, Liang Wu, Ying Gao, Shuo Cao, Yu Pan,
Phase transition and entropic force of de Sitter black hole in massive gravity,
Eur. Phys. J. C {\bf 81}, 42 (2021).

\bibitem{Odintsov2021}
S. Nojiri, S.D. Odintsov and V. Faraoni,
Area-law versus Renyi and Tsallis black hole entropies,
Phys. Rev. D {\bf 104}, 084030 (2021).

\bibitem{Shankaranarayanan2003}
S. Shankaranarayanan, 
Temperature and entropy of Schwarzschild-de Sitter space-time,
Phys. Rev. D {\bf 67}, 084026 (2003).

\bibitem{Singha2021}
C. Singha,
Thermodynamics of multi-horizon spacetimes,
General Relativity and Gravitation {\bf 54}, 38 (2022).
arXiv:2108.11704.

\bibitem{Azarnia2021}
S. Azarnia and S. Sedigheh Hashemi,
Correlation of horizons and black hole thermodynamics,
arXiv:2111.08984.

\bibitem{Volovik2022c}
G.E. Volovik,
Painlev\'e-Gullstrand coordinates for Schwarzschild-de Sitter spacetime,
arXiv:2209.02698 [gr-qc].

\end{thebibliography}
\end{document}